\documentclass[11pt]{article}
\usepackage[colorlinks=true, citecolor=blue, urlcolor=blue, linkcolor=blue, breaklinks=true, pdfpagelabels=false]{hyperref}
\usepackage{cite}

\usepackage{hyperref}

\usepackage{amsmath,amsfonts,amssymb}
\usepackage{graphicx}
\usepackage{epsfig,epsf}
\usepackage{bm}
\usepackage{xcolor}

\def\bar {\overline}

\def\be {\begin{equation}}
\def\ee {\end{equation}}
\def\bea {\begin{eqnarray}}
\def\eea {\end{eqnarray}}

\def\barr{\begin{array}}
\def\earr{\end{array}}

\def\opcit(#1){ {\em op. cit.}, #1}

\def\issue(#1,#2,#3){#1, #2 (#3)} % AIP format
\def\equationautorefname~#1\null{Eq.\,(#1)\null}
\def\pageautorefname\nobreakspace{p.}

\makeatletter\renewcommand{\p@subsection}{\thesection.}\makeatother%
\begin{document}

\renewcommand*{\thefootnote}{\fnsymbol{footnote}}

%%%%%%%%%%%%%%%%%%%%%%%%%%%%%%%%%%%%%%%%%%%%%%%%%%%%%%%

\begin{center}
{\Large\bf{Model Independent Effective Dimension Six Operators And Scattering Unitarity}}

%%%%%%%%%%%%%%%%%%%%%%%%%%%%%%%%%%%%%%%%%%%%%%%%%%%%%%%%

\vspace{5mm}

{\bf Swagata Ghosh}$^{a,b}$\footnote{swgtghsh54@gmail.com, speaker at $26^{th}$ DAE-BRNS HEP Symposium, December 19-23, 2024, BHU, India.}
{\bf Rashidul Islam}$^{c}$\footnote{islam.rashid@gmail.com}

\vspace{3mm}
{\em{${}^a$\ \ \ Department of Physics, Indian Institute of Technology Kharagpur, Kharagpur 721302, India.
}}

{\em{${}^b$\ \ \ Department of Physics, University of Calcutta, 92 Acharya Prafulla Chandra Road, Kolkata 700009, India.}}

{\em{${}^c$\ \ \ Department of Physics, Mathabhanga College, Cooch Behar, West Bengal, India.}}

\end{center}

%%%%%%%%%%%%%%%%%%%%%%%%%%%%%%%%%%%%%%%%%%%%%%%%%%%%%%%
\begin{abstract}

The effective field theory containing higher dimensional operators violates the unitarity of the $2 \rightarrow 2$ scattering processes in the Standard Model. 
This unitarity violation depends on the values of the Wilson coefficients corresponding to a higher dimensional operator. 
We showed that even a small values of some of the Wilson coefficients lead to the unitarity violation at the LHC centre of mass energies. 
Considering the final states as the gauge bosons in the longitudinal mode, the scalars, and the $t\overline{t}$, we showed that the unitarity violation in $WW \rightarrow WW$, $WW \rightarrow ZZ$, and $ZZ \rightarrow hh$ scattering amplitudes sets bounds on the Wilson coefficients of the dimension six effective operators. 

\end{abstract}

%\small{PACS no.: {12.60.Fr, 14.80.Ec}}

%\maketitle

\setcounter{footnote}{0}
\renewcommand*{\thefootnote}{\arabic{footnote}}

%%%%%%%%%%%%%%%%%%%%%%
\section{The effective Lagrangian and scattering amplitude}
We consider the set of dimension six operators of the Strongly Interacting Light Higgs (SILH) basis \cite{0703164,1303.3876} with the $i^{th}$ Wilson coefficient (WC) $\overline{c}_i$ corresponding to the $i^{th}$ operator $O_i$ coming with the cut-off scale $\Lambda$. 
The relevant bosonic and fermionic operators are given by, 
\begin{eqnarray}
 &&{\cal L}_{\rm boson} = \frac{1}{\Lambda^2}\sum_i \overline c_i O_i =
    \frac{\overline c_H}{\Lambda^2} \partial^\mu(\Phi^\dagger \Phi) \partial_\mu(\Phi^\dagger \Phi)
   + 
   \frac{\overline c_T}{\Lambda^2} ( \Phi^\dagger \overleftrightarrow{D})^\mu \Phi )( \Phi^\dagger \overleftrightarrow D^\mu \Phi ) 
  - 
  \frac{\overline c_6 \lambda}{\Lambda^2} (\Phi^\dagger \Phi)^3 \nonumber \\ &&
  + 
  \frac{i g \overline c_W}{2\Lambda^2}
     \ ( \Phi^\dagger \tau^i \overleftrightarrow D^\mu \Phi )
      \ ( D^\nu W_{\mu\nu} )^i
  + 
  \frac{ig' \overline c_B}{2\Lambda^2}
     \ ( \Phi^\dagger \overleftrightarrow D^\mu \Phi )
      \ ( \partial^\nu B_{\mu\nu} ) 
  + 
  \frac{ig\overline c_{HW}}{2\Lambda^2}
     \ (D^\mu \Phi)^\dagger \tau^i (D^\nu \Phi)
      \ W^i_{\mu\nu} \nonumber \\ &&
  + 
  \frac{ig'\overline c_{HB}}{2\Lambda^2}
     \ (D^\mu \Phi)^\dagger (D^\nu \Phi) \ B_{\mu\nu} 
  + 
  \frac{{g'}^2 \overline c_\gamma}{\Lambda^2} \ (\Phi^\dagger \Phi) \ B_{\mu\nu} B^{\mu\nu}
  + 
  \frac{g_s^2 \overline c_g}{\Lambda^2} \ (\Phi^\dagger \Phi) \ G^a_{\mu\nu} G^{a\mu\nu}  \nonumber \\ &&
  + 
  \frac{g^3 \overline c_{3W}}{\Lambda^2} \epsilon_{ijk}
  W^{i\nu}_\mu W^{j\alpha}_\nu W^{k\mu}_{\alpha}\,,
\end{eqnarray}
\begin{eqnarray}
 &&{\cal L}_{\rm fermion} = \frac{1}{\Lambda^2}\sum_j \overline c_j O_j =
 \left(
 \frac{\overline{c}_u}{\Lambda^2}  y_u \overline{Q}_L u_R \Phi^c \Phi^\dagger\Phi
 + 
 \frac{i\overline{c}_{Hud}}{\Lambda^2}  (\overline u_R \gamma^\mu d_R) (\Phi^{c\dagger} \overleftrightarrow
 D_\mu \Phi) 
 +{\rm h.c.}
 \right) \nonumber \\ &&
 + 
 \frac{i\overline{c}_{Hq}}{\Lambda^2} (\overline Q_L \gamma^\mu Q_L) (\Phi^\dagger \overleftrightarrow D_\mu \Phi) 
 + 
 \frac{i\overline{c}'_{Hq}}{\Lambda^2} (\overline Q_L \tau^i \gamma^\mu Q_L) (\Phi^\dagger \tau^i
 \overleftrightarrow D_\mu \Phi)  
 + 
 \frac{i\overline{c}_{Hu}}{\Lambda^2} (\overline u_R \gamma^\mu u_R) (\Phi^\dagger \overleftrightarrow D_\mu \Phi) \nonumber \\ &&
 + 
 \frac{i\overline{c}'_{Hd}}{\Lambda^2} (\overline d_R \tau^i \gamma^\mu d_R) (\Phi^\dagger \tau^i
 \overleftrightarrow D_\mu \Phi) 
 + 
 \frac{g' \overline{c}_{uB}}{\Lambda^2} y_u \overline Q_L \Phi^c \sigma_{\mu\nu} u_R B^{\mu\nu}
 + 
 \frac{g \overline{c}_{uW}}{\Lambda^2} y_u \overline Q_L \tau^i \Phi^c \sigma_{\mu\nu} u_R W^{i\mu\nu} \nonumber \\ &&
 + 
 \frac{g' \overline{c}_{dB}}{\Lambda^2} y_d \overline Q_L \Phi \sigma_{\mu\nu} d_R B^{\mu\nu}
 + 
 \frac{g \overline{c}_{dW}}{\Lambda^2} y_d \overline Q_L \tau^i \Phi \sigma_{\mu\nu} d_R W^{i\mu\nu}\,.
\end{eqnarray}
Here, $\Phi^\dag {\overleftrightarrow D}_\mu \Phi = \Phi^\dag D^\mu \Phi - D_\mu\Phi^\dag \Phi$, $\tau_i$s are the Pauli matrices, $g'$, $g$, and $g_s$ are the U(1)$_Y$, SU(2)$_L$, and SU(3)$_c$ gauge couplings respectively. 
The operators relating to the gluons and leptons are not considered in this work. 

The amplitude $A$ of any $2 \rightarrow 2$ scattering process can be expressed in terms of the partial waves $a_l$ as,
$   
A = 16\pi \sum_{\ell=0}^\infty \, (2\ell+1) \, P_\ell(\cos\theta)\, a_\ell.
$
Using the optical theorem for zero scattering angle $(\theta = 0)$, we get, 
$
 |a_l|^2 = {~\rm{Im}}\,\, a_l \Rightarrow {~\rm Re}\,\, a_l \le \frac12. 
$
We are interested in $0^{th}$ order partial waves $a_0$ only. 
\section{Redefinition of fields and masses}
To express the kinetic terms in their canonical form, the fields can be redefined as follows considering $\overline{c}_iv^2/\Lambda^2 \ll 1$, 
\begin{eqnarray}
 &&h \to  h \left[ 1 - \dfrac{\bar c_H}{\Lambda^2} v^2 \right]\,,\,\,
 G^a_\mu \to  G^a_\mu \left[1 + \dfrac{\bar c_G}{\Lambda^2} g^2_s v^2\right]\,,\,\,
 W^\pm_\mu \to W^\pm_\mu \left[1 + \dfrac{\bar c_W}{\Lambda^2} \dfrac{g^2}{4} v^2 \right]\,,
 \nonumber \\&&
 Z_\mu \to Z_\mu\left[1 + \dfrac{\bar c_\gamma}{\Lambda^2} {g'}^2 \sin^2\theta_W v^2
   + \dfrac{\bar c_W}{\Lambda^2} \dfrac{g^2}{4} v^2+ \dfrac{\bar c_B}{\Lambda^2}
   \dfrac{g^{\prime2}}{4} v^2\right]\,,\nonumber \\&&
 A_\mu \to A_\mu\left[1 + \dfrac{\bar c_\gamma}{\Lambda^2} g^2 \sin^2\theta_W v^2
  \right] + Z_\mu \dfrac{\bar c_W - \bar c_B - 8 \bar c_\gamma \sin^2\theta_W}{\Lambda^2}
   \dfrac{g g'}{4} v^2\,.
\end{eqnarray}
Eventually, the particle masses are also modified to 
\begin{eqnarray}
 &&m^2_h = \left[1 - 2 v^2 \dfrac{\bar c_H}{\Lambda^2}\right]
  (3 \lambda v^2 - \mu^2) + \dfrac{15}{4} \lambda v^4 \dfrac{\bar c_6}{\Lambda^2} \equiv 3 \lambda v'^2 - \mu'^2\,,\nonumber \\&&
 m^2_Z = \dfrac{g^2 {v'}^2}{4 \cos^2\theta_W}\left[
  1 + g^2 {v'}^2 \dfrac{\bar c_W}{2 \Lambda^2} + {g'}^2 {v'}^2 \dfrac{\bar c_B}{2 \Lambda^2}
  + 2 {g'}^2 {v'}^2 \sin^2\theta_W \dfrac{\bar c_\gamma}{\Lambda^2}
  - 2 {v'}^2 \dfrac{\bar c_T}{\Lambda^2}\right]\,, \nonumber \\ &&
 m^2_W = \dfrac{g^2 {v'}^2}{4} \left[1 + g^2 {v'}^2 \dfrac{\bar c_W}{2 \Lambda^2}\right]\,,
 \quad
 {\rm with}\,\,
 v'^2 = v^2 - \frac{v^4}{\Lambda^2}\left[2\overline{c}_H - \frac54\overline{c}_6\right]\,.
\end{eqnarray}
Here, $v$ and $v'$ are the Standard Model and redefined Higgs vevs respectively. 
\section{Bounds on the Wilson coefficients}
We set the cutoff scale and the parton-level energy respectively at $\Lambda = 1$ TeV, and $\sqrt{s} = 2$ TeV. 
We implement the effective Lagrangian in FeynRules \cite{feynrules} to generate FeynArts model files to calculate the helicity amplitudes using FeynArts/FormCalc \cite{feynarts,formcalc}. 
 \begin{figure}
  \begin{center}
  \includegraphics[width=5.9cm]{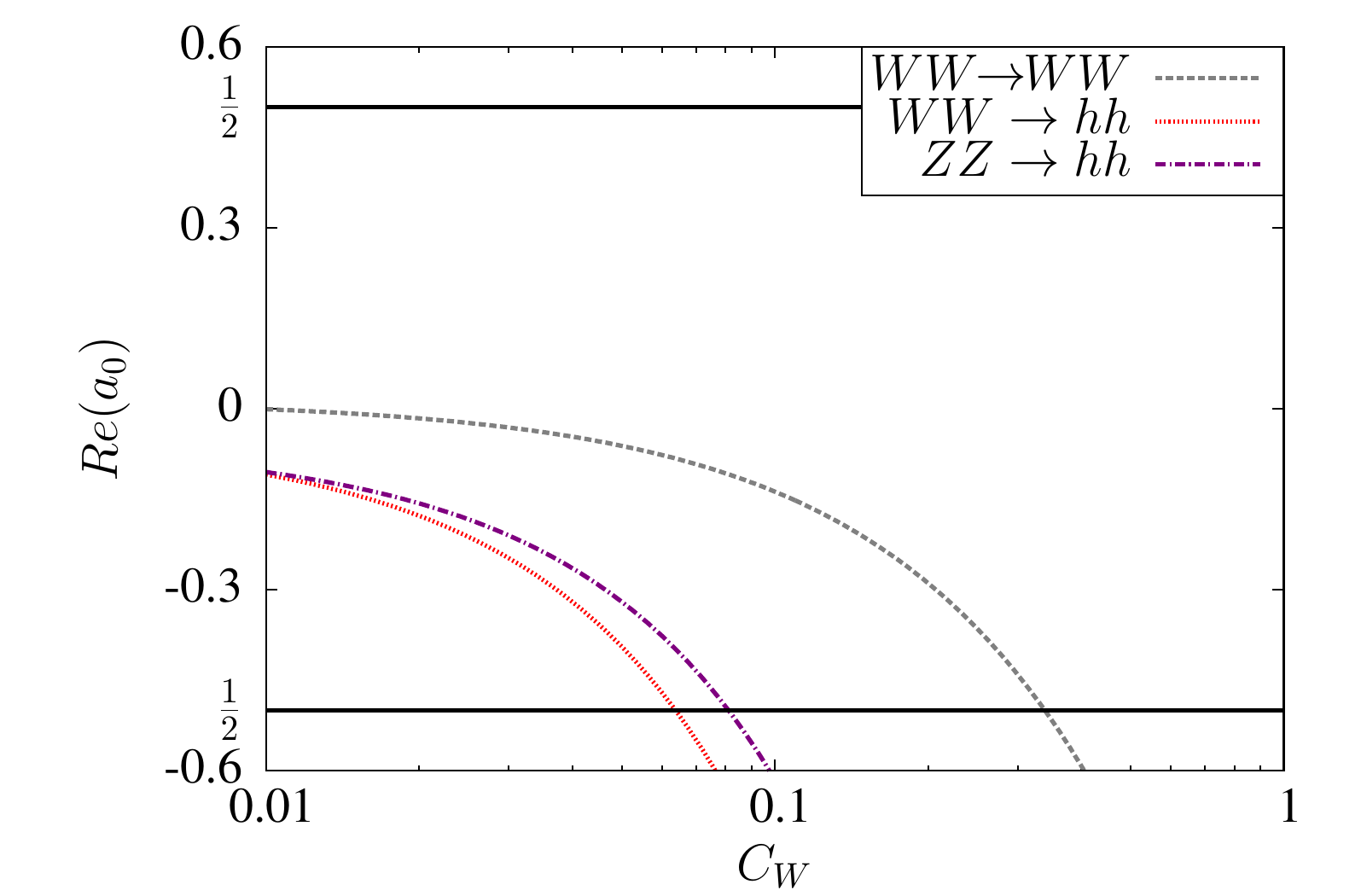} \ \
  \includegraphics[width=5.9cm]{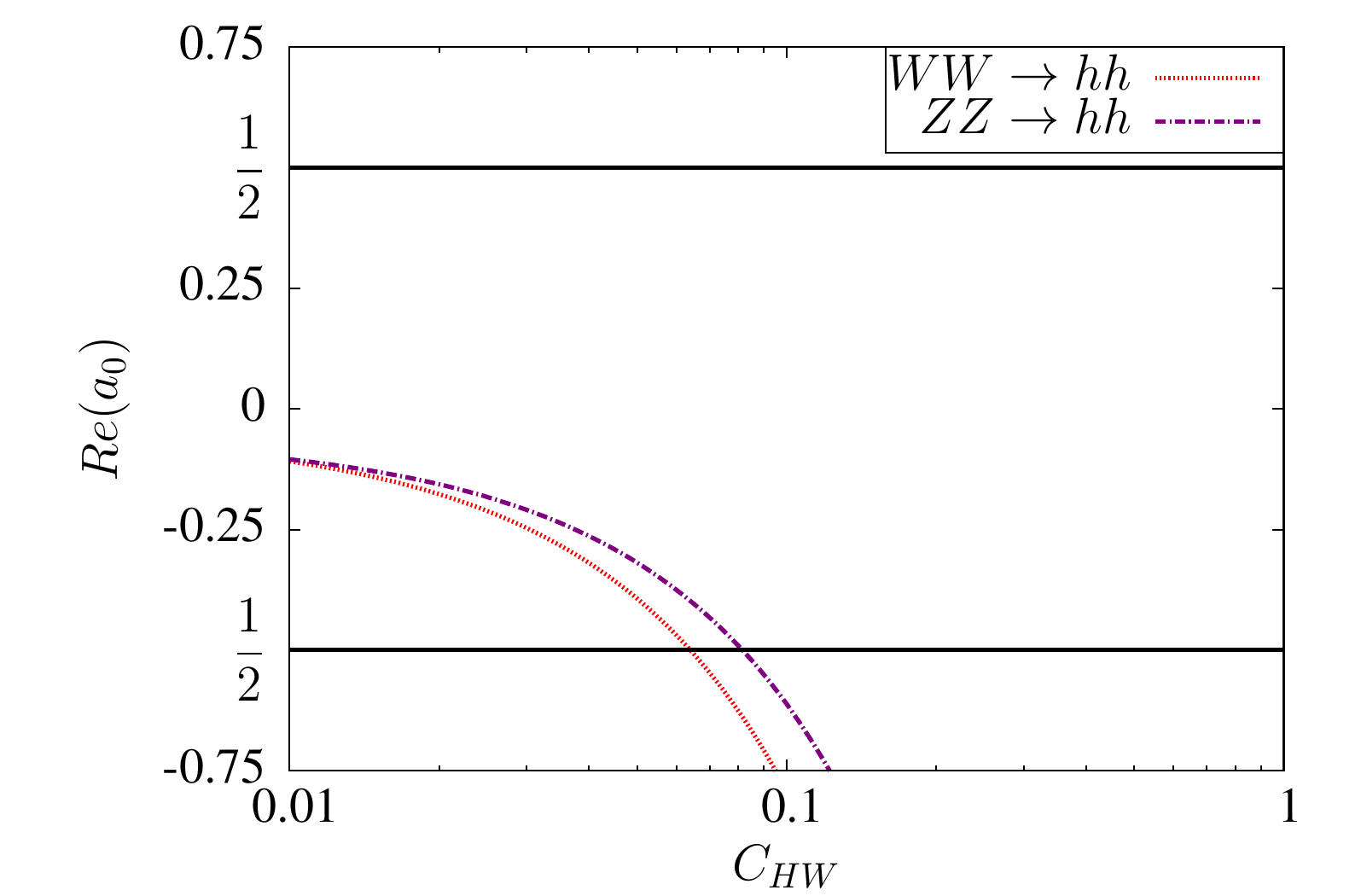}\\
  \includegraphics[width=5.9cm]{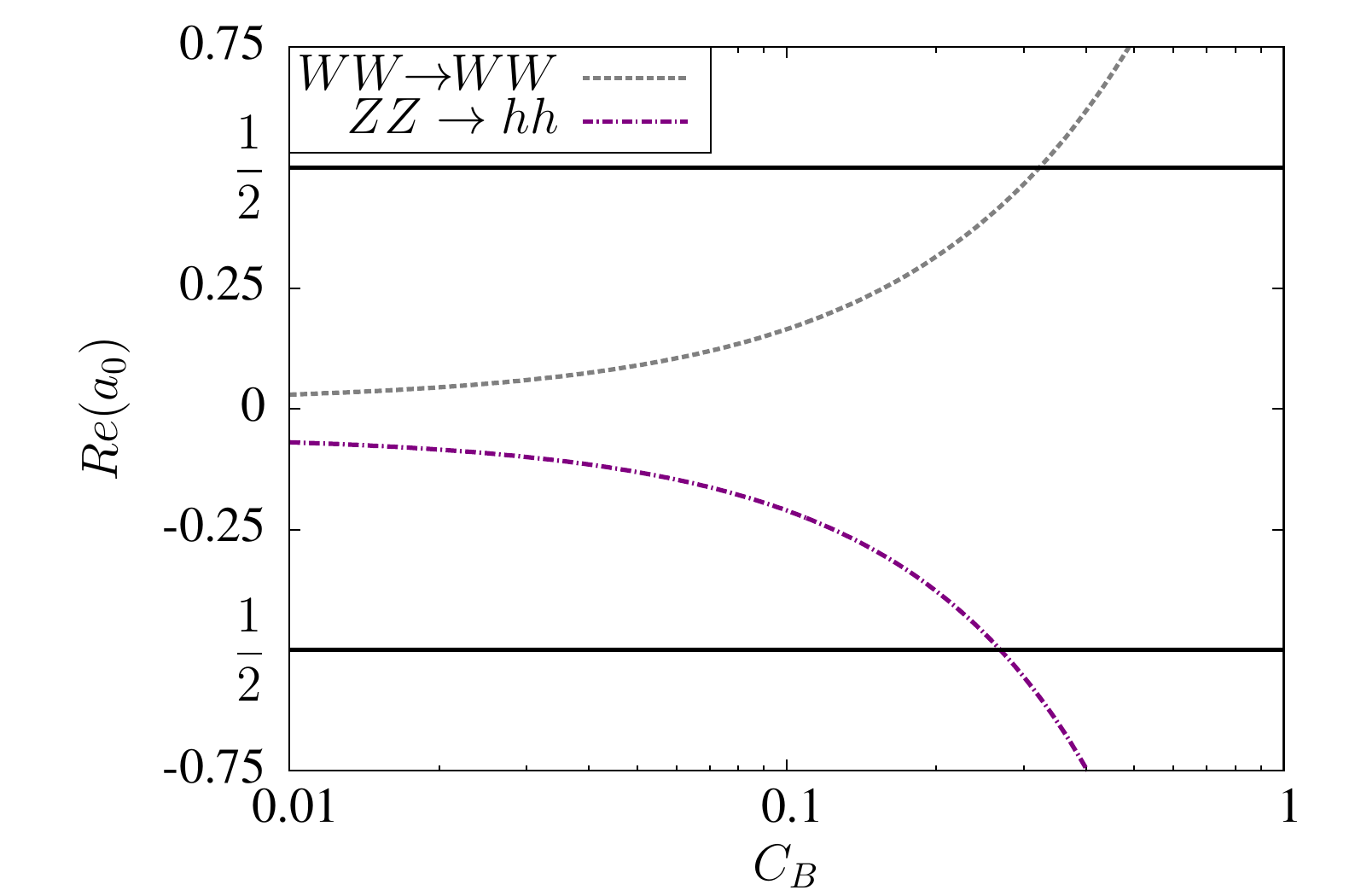}\ \
  \includegraphics[width=5.9cm]{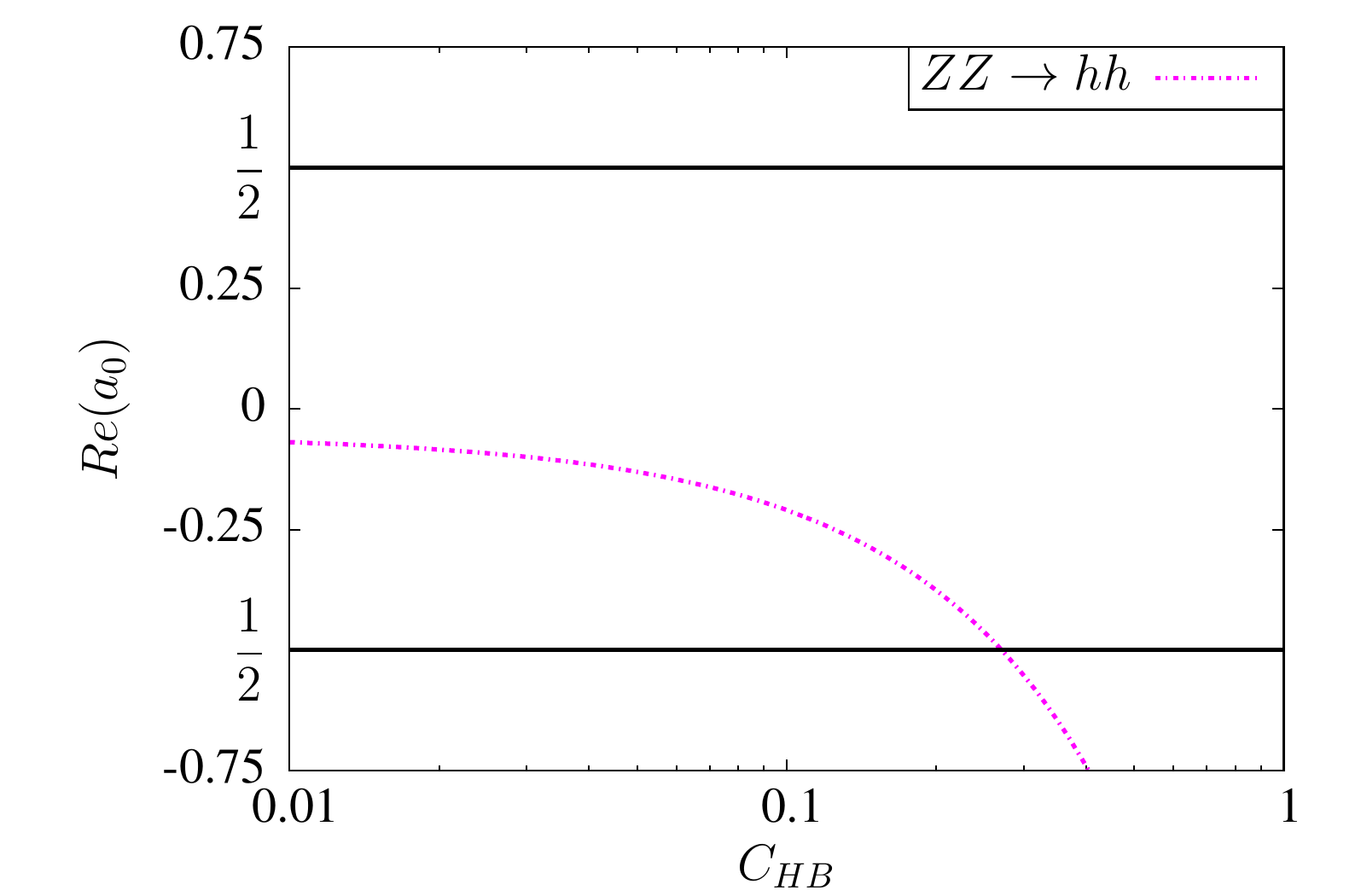}\\
  \includegraphics[width=5.9cm]{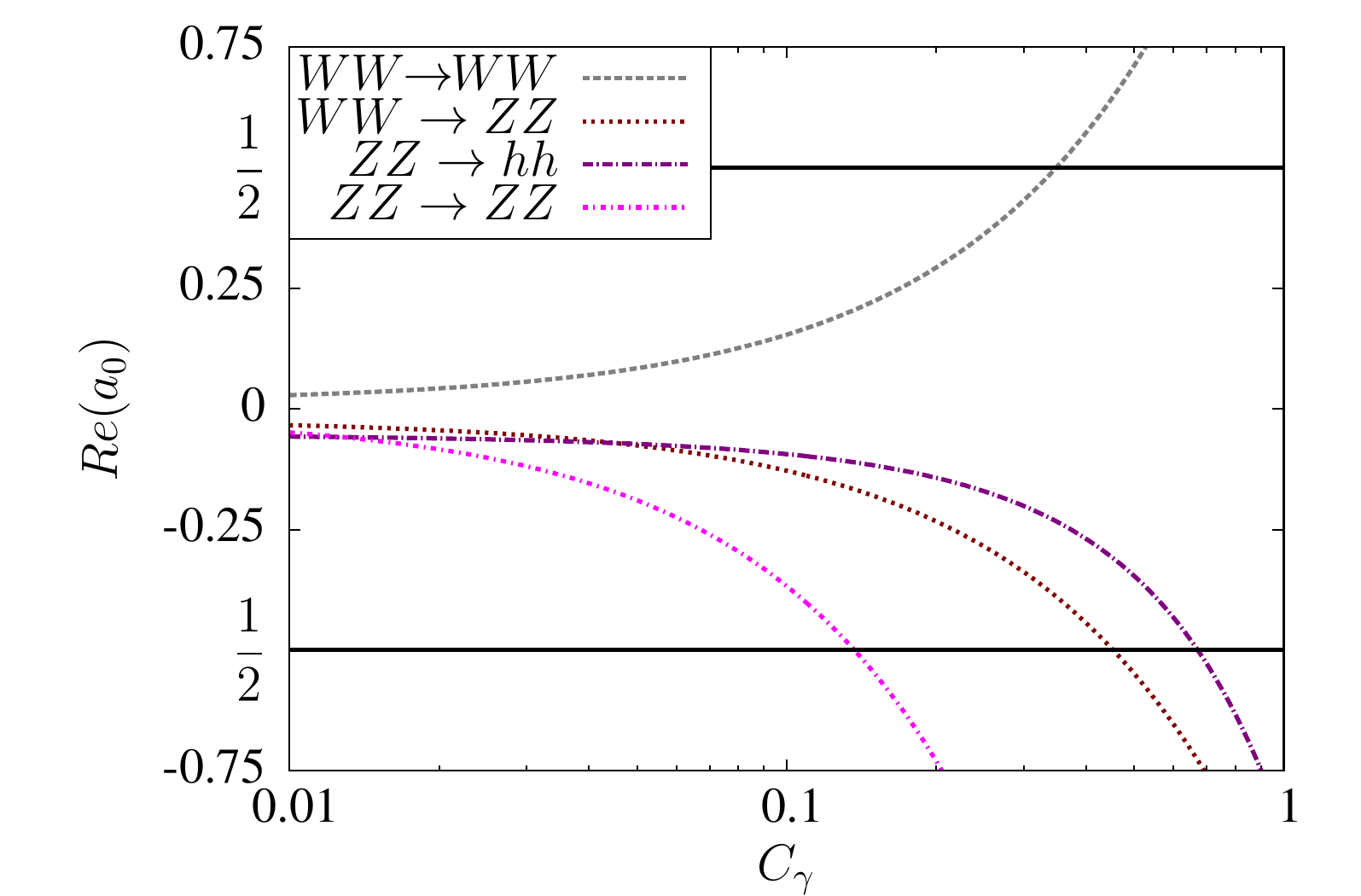}\ \
  \includegraphics[width=5.9cm]{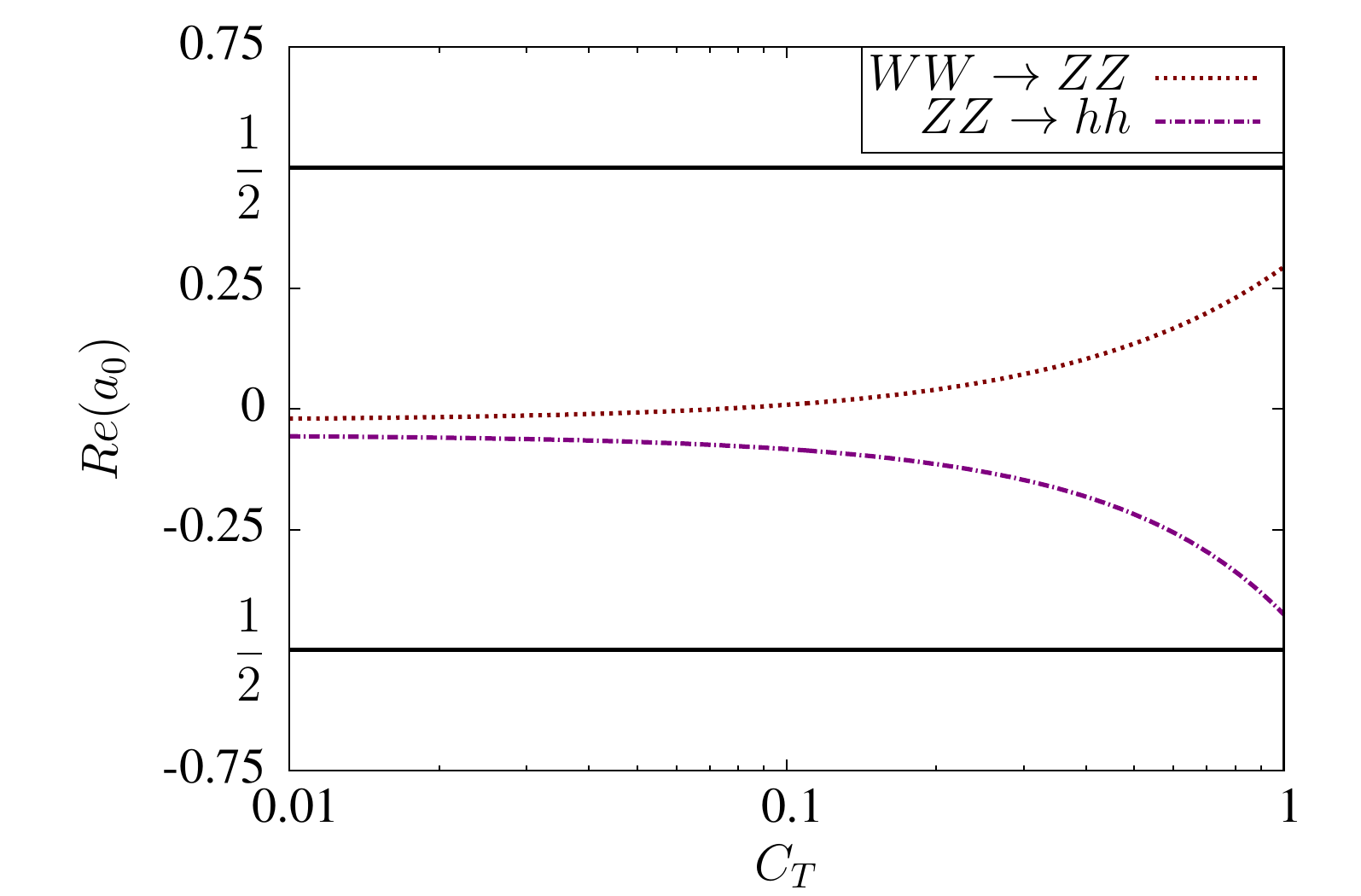}\\
  \caption{The unitarity limits on the effective Wilson coefficients, where they violate the bound $|{\rm Re}\, a_0| \le \frac12$. }
  \end{center}
 \end{figure}
 \begin{table}
 \small
  \begin{center}
   \begin{tabular}{|| c || c | c | c | c | c | c | c | c | c ||}
    \hline\hline
%    &&&&&&&&&\\
    PROCESS     & $O_6$   & $O_T$   & $O_H$   & $O_\gamma$  & $O_W$   & $O_{HW}$ & $O_B$   & $O_{HB}$ & $O_{3W}$ \\
%    &&&&&&&&&\\ 
    \hline\hline
    $WW\to WW$  &         &         & $\otimes$ & $\otimes$ & $\surd$ $\otimes$ & $\surd$  &  $\otimes$   & $\surd$ & $\surd$   \\ \hline
    $WW\to ZZ$  &         &  $\surd$  &  $\otimes$ & $\surd$ $\otimes$  & $\surd$ $\otimes$ & $\surd$
    & $\otimes$  & $\surd$ & $\surd$  \\ \hline
    $ZZ\to ZZ$  &         & $\surd$ & $\otimes$ & $\surd$ $\otimes$ & $\otimes$ & $\surd$  & $\otimes$
    & $\surd$ &   \\ \hline
    %%%%%%%
    $WW\to hh$  & $\surd$ &         & $\surd$ $\otimes$ &         & $\surd$ $\otimes$
    & $\surd$  &         &   $\surd$ & \\ \hline
    $ZZ\to hh$  & $\surd$ & $\surd$ & $\surd$ $\otimes$ & $\surd$ $\otimes$ &
    $\surd$ $\otimes$ & $\surd$  & $\surd$ $\otimes$ & $\surd$  &  \\ \hline
    $WW \to t\bar{t}$ &      &            &  $\otimes$  & $\otimes$  &$\surd$ $\otimes$& $\surd$  & $\otimes$   &
     & $\surd$    \\ \hline
    $ZZ \to t\bar{t}$ &      & $\surd$  & $\otimes$  & $\surd$ $\otimes$ & $\otimes$ & $\surd$ &
    $\otimes$ & $\surd$ & \\ \hline
%    $hh\to hh$  & $\surd$ &         & $\surd$ &         &         &          &         &           \\ \hline
   \end{tabular}
   \caption{\small{Dimension-6 operators affecting the bosonic and fermionic
   scatterings. The entries marked with $\surd$
   are affected by the modification of the SM vertices. The entries marked with $\otimes$ are affected by the
   wavefunction normalization.}}
  \end{center}
  \label{tab:1}
 \end{table}
The Table \ref{tab:1} shows which operators contribute to which scattering processes. 
As we take $\Lambda = 1$ TeV, the WCs are scaled with $\Lambda^2$, and we express $\overline{c}_i/\Lambda^2 = C_i$. 
We took one operator to be non-zero at a time and drop $O_6,\, O_H,\, O_{3W}$ as they do not violate the unitarity. 
$WW\to hh$ provides bounds on $\overline{C}_W$ and $\overline{C}_{HW}$ both at $0.06$, whereas 
$ZZ\to hh$ provides bounds on $\overline{C}_B$ and $\overline{C}_{HB}$ both at $0.27$. 
We have bounds on $\overline{C}_{\gamma}$ at $0.14$ and $\overline{C}_T$ at $1.2$ from the processes $ZZ\to ZZ$ and $ZZ\to hh$ respectively. 
A more detailed study can be found in \cite{Ghosh:2017coz}.

\section{Conclusion}
The introduction of the dimension six operators causes unitarity violation of some of the scattering amplitudes. 
We consider here the dimension six operators of the SILH basis. 
Some of the operators affect the canonical kinetic terms causing the modification of the fields and their masses and hence indirectly affect the vertices as shown in the Table \ref{tab:1}. 
The Wilson coefficients come with a cut-off scale $\Lambda$, but always scale with $\Lambda^2$. 
Therefore, we set $\Lambda = 1$ TeV to show the bounds on the WCs. 
All the bounds are shown for $\sqrt{s} = 2$ TeV, the typical parton-level energy for the LHC. 
We consider one non-zero operator at a time in this work. 
Using optical theorem for zero scattering angle and zeroth partial wave $a_0$ we showed the bounds on different Wilson coefficients when the terms linear in $c_i/\Lambda^2$ are sufficient. 
%
% ---- Bibliography ----
%

\end{document}